# Mid-infrared quantum cascade laser frequency combs with a microstrip-like line waveguide geometry

Filippos Kapsalidis,[1, a)] Barbara Schneider,[1] Matthew Singleton,[1] Mathieu Bertrand,[1] Emilio Gini,[2] Mattias Beck,[1] and Jérôme Faist[1]

[1)] *Institute for Quantum Electronics, ETH Zürich, Switzerland*
[2)] *FIRST - Center for Micro- and Nanoscience, ETH Zürich, Switzerland*

(Dated: 16 December 2020)

In this work, a design for a mid-infrared quantum cascade laser (QCL) frequency comb source that enhances the high frequency response and the comb characteristics of the device is presented . A state-of-the-art active region (AR), grown on a heavily n-doped InP:Si substrate, was processed into a buried heterostructure with a microstrip-like line waveguide. As a result, the repetition rate frequency $f_{rep}$, around 11.09 GHz, can be locked to an injected narrow-linewidth radio frequency (RF) signal, over a range of more than 200 kHz with -10 dBm of injected power, which outperforms normal buried heterostructure schemes by an order of magnitude. Moreover, under RF injection at powers higher than 20 dBm, the lasing spectrum is flattened and significantly broadened, from 24 $cm^{-1}$ to 65 $cm^{-1}$ in bandwidth, while at the same time the coherence of the comb is maintained and verified.

## I. INTRODUCTION

Quantum cascade lasers (QCLs) are unipolar semiconductor devices, based on intersubband transitions in coupled quantum wells and barriers systems, which have become the dominant light source in the mid-infrared part of the electromagnetic spectrum[1] and are rapidly becoming serious contenders for the terahertz (THz)[2,3]. These devices are monolithic, electrically pumped, and they naturally operate as frequency combs in both the mid-infrared[4] and THz[5], making them prized devices for coherent spectroscopy[6–9] in those spectral regions. The characteristics of a QCL device operating as a frequency comb are a broad and phase-locked optical spectrum of equidistant modes, together with a strong narrow radio frequency (RF) beatnote generated by the beating of these optical modes[4].

One unique property of QCLs is their ultrafast carrier relaxation lifetimes, which are in the order of picoseconds (ps), mainly influenced by phonon assisted intersubband scattering[10]. This means that carrier populations can track the intracavity beating between the optical modes, which, combined with an intrinsic cut-off in the tens of GHz[10,11], permits the electronic beat signal at the repetition rate frequency $f_{rep}$ to be observed on the bias line. This logic applies vice versa, allowing a coupling between the optical modes and an injected RF signal[12–14].

High frequency modulation and injection locking is extremely interesting for the field of QCL frequency combs and dual-comb spectroscopy. It has been experimentally proven that by injecting a microwave signal at the cavity $f_{rep}$ it is possible to further stabilize the comb[15]. It is also an important element for active mode locking of QCLs, demonstrated in the THz range[13,16,17] but also in the mid-infrared for interband cascade lasers[18] and QCLs[19].

In this work, we are interested in further exploring the response and prospects of controlling the QCL comb state when the source laser geometry has been optimized for external microwave modulation. One of the key factors for efficiently injecting a fast external modulation signal, and using it to dynamically control the QCL comb, is ultimately the device package, that is, the device geometry, and its biasing network, used to inject the external signals. In this aspect, we designed a scheme for a mid-infrared QCL waveguide, similar to a microwave microstrip, where a dielectric is sandwiched between two metallic layers acting as a microwaves waveguide. That approach, commonly known in the QCL community as "double metal" configuration, while extensively used in the THz[20], for mid-infrared QCLs has shown limited performance and devices operated only at cryogenic temperatures[21–23], as having metallic layers in proximity to the active region introduces a great deal of optical losses[24].

## II. DEVICE CONCEPT, DESIGN AND FABRICATION

In contrast to previous double-metal designs, instead of using metallic layers for the top and bottom contacts, and wafer bonding the device on a substrate, a different approach is taken: the device waveguide is monolithically fabricated on a very heavily n-doped InP:Si substrate, together with a likewise heavily n-doped upper cladding, designed to replace both metallic microstrips, thereby controlling and minimizing the optical losses. The epitaxial topside of the devices is reduced in size, aiming to decrease parasitic capacitance, to extend the cut-off frequency and therefore improve the electrical response to RF signals, and to enhance the coupling between the optical comb formation and the carrier transport[25]. A schematic of the conceived device is illustrated in Fig. 1(a).

The device substrate is a heavily n-doped InP:Si, with a doping concentration of $\approx 8 \times 10^{18} cm^{-3}$. In order to avoid high optical losses induced by the many free carriers, the QCL active region (AR) is not directly grown on the substrate. Instead, firstly a buffer of two lower doped ($2 \times 10^{16} - 1 \times 10^{17} cm^{-3}$) InP:Si layers is grown by metal-organic vapor epitaxy (MOVPE). After the InP buffer layer, the QCL active region structure is grown by means of molecu-

---
a)Electronic mail: fkapsali@phys.ethz.ch

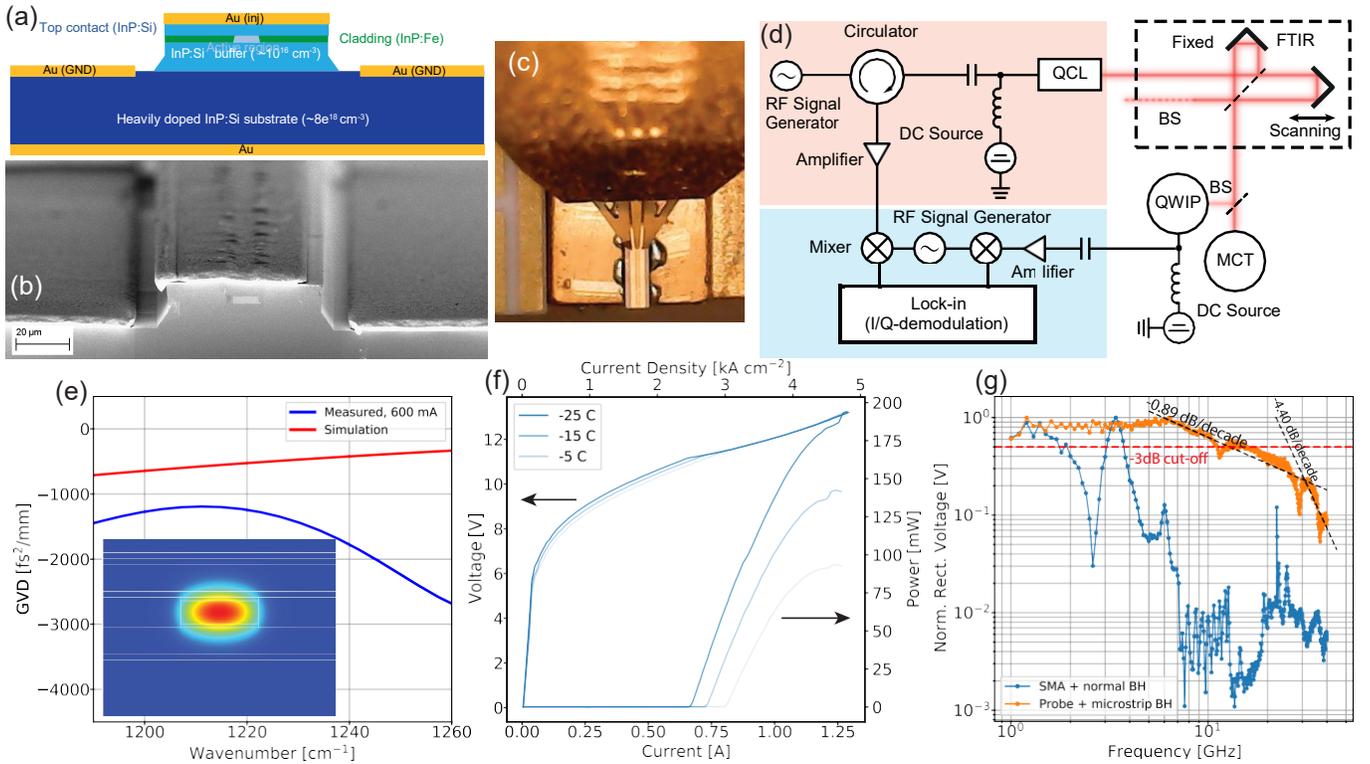

FIG. 1. (a) Schematic of a QCL frequency comb with a microstrip-like line waveguide geometry. (b) SEM micrograph of a finalized fabricated device. (c) The |Z|-probe in contact with the microstrip-like QCL. (d) The experimental setup: the RF signal generation and modulation is in the red shaded area and the coherent detection in the blue shaded area. The optical signal from the QCL is detected by both an MCT (Mercury Cadmium Telluride detector) and a QWIP (Quantum Well Infrared Photodetector) behind an interferometer in a folded Mach-Zehnder configuration. The electrical beatnote is extracted via a bias-tee. A circulator is used to enable simultaneous injection and detection. (e) The simulated (red) and measured (blue) curves of the group velocity dispersion (GVD) around the relevant spectral span. Inset: the simulated mode at 1220 $cm^{-1}$. (f) LIV characteristic curves at -25, -15 and -5°C. (g) Normalized rectification voltage for two kinds of devices: normal buried heterostructure frequency comb QCL (blue curve), microstrip-like line waveguide QCL optimized for RF injection (orange curve).

lar beam epitaxy (MBE). It consists of a strain compensated In$_{0.592}$Ga$_{0.408}$As / In$_{0.36}$Al$_{0.65}$As heterogeneous stack of two bound-to-continuum active regions[26], covering two different wavelength regions at 8.02 $\mu m$ (1247 $cm^{-1}$) and 8.45 $\mu m$ (1183 $cm^{-1}$).

The rest of the n-doped InP:Si top cladding is grown additionally by MOVPE. The layer sequence can be found in Table S1 in the supplementary document. The strategy adopted for the top cladding was to gradually increase the n-doping concentration further away from the QCL active region, aiming to increase overlap with the optical mode, to keep optical losses to a minimum and the group velocity dispersion (GVD) flat and close to zero[27,28]. A simulation of the expected GVD of the fundamental optical mode is shown in Fig.1 (e, red curve).

The grown material is then fabricated into QCL buried heterostructure devices[29,30], with AR widths between 6 and 8 $\mu m$. The devices are then further processed into a microstrip line waveguide configuration with parallel ground electrodes: the sides of the laser ridge are etched down by$\approx$16 $\mu m$ to reach the heavily doped InP:Si substrate, leaving a 40 $\mu m$ wide topside central contact. In the etched trenches, metallic layers are also deposited, to work as ground contacts, each at a lateral distance of 15 $\mu m$ from the central metallic top contact. Finally, a 4 $\mu m$ thick layer of gold is electroplated on all of the metallic contacts, to improve heat extraction, device robustness and contact with the probe electrodes. See in Fig. 1(b) a SEM micrograph of a finished device.

The devices are lapped down to$\approx$190 $\mu m$ and a metallic layer is deposited to the bottom side, to help with soldering and heat extraction. The devices are finally soldered epitaxial side up on copper mounts.

As a reference, QCL combs with the same active material were fabricated with a normal buried heterostructure configuration[26]. Those devices were mounted epitaxial side down on AlN substrates. For the electrical injection scheme, SMA connectors with wire bonding were used, as is the standard method in the QCL comb and dual-comb spectroscopy community.

## III. EXPERIMENTAL METHODS - CHARACTERIZATION

In order to fully exploit the high frequency response of the microstrip-like devices, the use of wire bonds for biasing the devices is altogether avoided, as they introduce additional inductive contributions, and thus limit the modulation bandwidth at higher RF injection frequencies. Instead, the DC-current and RF injection is realized by a three terminal RF probe (FormFactor |Z| Probe Power, model Z040-P-GSG-250), compatible with the device geometry (Fig. 1 (b,c)). A bias tee separates the DC and RF components, and a circulator (-20 dB isolation) allows for simultaneous injection and monitoring of the electrical beatnote.

For spectral analysis in the RF and mid-infrared, a dedicated setup was built as shown in the schematic drawing in Fig. 1 (d). The optical output from the laser is collimated into a home-made Fourier Transform Infrared Spectrometer (FTIR) mounted in a folded Mach-Zehnder configuration, with a resolution down to 0.015 cm$^{-1}$. The optical output of the FTIR is split and shined onto a sensitive Mercury Cadmium Telluride (MCT) detector to collect the DC autocorrelation traces, and a fast Quantum Well Infrared Detector (QWIP), which is used for measurement of RF correlation traces.

### A. Device Performance

The light-current-voltage (LIV) characteristics of a 4 mm long and 6.5 μm wide device, operating in CW at -25, -15 and -5 °C are shown on Fig. 1 (f) with a maximum optical power of 200 mW at -25 °C. For higher operating temperatures the performance rapidly deteriorates. This is due to the large width of the active region, the reduced size of the top contact, and the epitaxial side up mounting of the device, all of which reduce the thermal extraction efficiency. The heatsink was fixed at -25°C for all following experiments.

The GVD of the device is measured below the lasing threshold, at 600 mA, using a Fourier transform of the interferogram of the amplified spontaneous emission[31]. The obtained values are shown in Fig. 1 (e) with the blue curve. The values of the dispersion are between -1000 to -2000 fs$^2$/mm around the spectral area of lasing. The discrepancy to the simulated GVD (red curve) is attributed to limitations of our model for accurately calculating the dispersion of the refractive index of the active region.

To assess the RF response of the microstrip-like devices at high modulation frequencies, a microwave rectification technique is used[25,32]. A lock-in amplifier measures the DC rectification while the device is driven simultaneously with a DC current source and an amplitude modulated (AM) signal whose carrier frequency is being swept across the whole frequency range of the RF source. The frequency chosen for the AM modulation signal is set at 50 kHz, its modulation depth at 10% and sweeping step at 100 MHz. The current density of both reference and microstrip-like lasers was set subthreshold at 2.2 kA/cm$^2$.

In Fig. 1 (g) we compare the response of a reference BH QCL comb (blue curve) and the microstrip-like QCL comb (orange curve). For the reference device there is a sharp reduction in the response at around 2 GHz modulation, as indicated by the 3 dB cut-off (red dashed line). The resonance peak at 3.5 GHz is attributed to cavity effect induced by the wire bond connection between the laser substrate and the contact pads. For the microstrip-like device, there is not such a sharp cut-off as with the reference. The trend of the curve after 6 GHz is showing a smooth decay, with a slope of -0.89 dB/decade. The 3dB limit is passed at 13 GHz, with the trend continuing until 30GHz, where the slope is -4.40 dB/decade. For this device, the rectification is no higher than one order of magnitude at 40 GHz, compared to three orders of magnitude of that of the reference device. This demonstrates the applicability of the microstrip-like line waveguide design and the probe for efficient RF injection and control of the device.

### B. Injection-locking and SWIFTS

Injection locking is a physical mechanism that occurs in systems of coupled harmonic oscillators. If the coupling between the two is strong enough, and oscillation frequencies are similar, then the primary oscillator can force the secondary into synchronisation; this means that properties of the primary, such as the oscillation frequency - and by extension, linewidth - are transferred onto the secondary. If instead the coupling is weak or the frequencies are further away from each other, injection pulling (of the secondary's frequency towards the primary's) is instead observed. The frequency range over which the locking between the primary and secondary oscillators is maintained, is given by Adler's equation[33]:

$$\Delta \omega_{locking} = \frac{2\omega_{free}}{Q} \sqrt{\frac{P_{inj}}{P_{free}}} \quad (1)$$

Here, $\omega_{free}$ is the frequency of the free running (secondary) oscillator, $P_{inj}$ and $P_{f\,ree}$ are the powers of the injected (primary) signal and the free running oscillator respectively, and Q the quality factor of the free running oscillator resonator.

In Fig. 2 (a) the behavior of the device during RF injection locking at an injection power of -11 dBm is reported. Here, we indeed observe the predicted behavior where at $|\omega_{inj} - \omega_{f\,ree}| \gg \Delta\omega_{locking}$ the position of the laser RF beatnote remains unperturbed but sidebands at integer multiples of $|\omega_{inj} - \omega_{f\,ree}|$ are generated. As $\omega_{inj}$ approaches $\omega_{f\,ree}$, the beatnote is pulled towards the injected signal and once $|\omega_{inj} - \omega_{f\,ree}| < \Delta\omega_{locking}$ the beatnote follows the injected signal until it leaves the locking range again. The locking range of the microstrip-like devices can stably reach more than 200 kHz.

Comparing the locking bandwidth for low injection powers of microstrip-like comb devices with that of reference buried heterostructure ones with a similar active region, as shown in Fig. 2 (b), there is more than an order of magnitude difference. It should be noted that here the values of the instrument are reported as the injection power. The solid lines represent fits of equation 1, which agree quite well with the recorded data. From the fit of the microstrip device, and since it is shown that for this injection frequency range the response of the device is enhanced, one can then reliably extract the values of the quality factor Q of the resonator. In this case, the extracted value is 7.42 ×10$^7$, same order of magnitude as when free-running

($f/\Delta f = 1.1 \times 10^7$, inset of Fig. 2(b)). The above demonstrate clearly that the microstrip-like line design has lead to an increased RF injection efficiency.

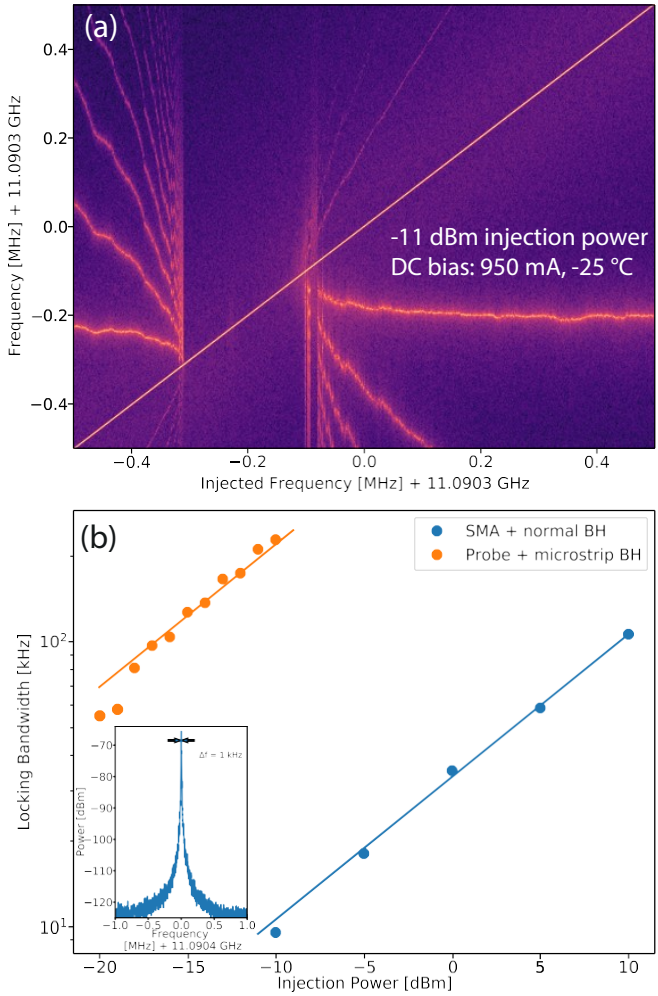

FIG. 2. (a) Electronic RF spectrum of current modulation inside the QCL cavity as a function of frequency, for -11 dBm RF generator output power. (b) Locking ranges of a microstrip-like line device (orange dots) at different RF generator output powers, compared to those for a reference buried heterostructure QCL comb mounted epitaxial side down and contacted with wire bonds (blue dots). Inset: Free-running RF beatnote of the microstrip device.

We assess the comb properties and coherence of the microstrip-like device output via the shifted wave interference Fourier transform spectroscopy (SWIFTS)[5,34]. Additionally, the time dependent instantaneous intensity and frequency of the coherent part of the laser output are reconstructed[5,15,34].

In Fig. 3 (a), the SWIFTS spectra, and the respective spectrum products, are shown for injection powers between -5 and 35 dBm in steps of 10 dBm, along with the free-running case. They have been normalized such that the relative powers of the two traces scale the same between measurements. A spectral bandwidth broadening from the free-running width of 24 $cm^{-1}$ to a width of 65 $cm^{-1}$ is observed along with a flattening of the amplitude distribution, for increasing RF injec-tion powers, without a clear loss in coherence, expect for the case of 35 dBm injection, where part of the lasing spectrum becomes unstable at the transition from the main lobe to a smaller shoulder towards higher frequencies. Similar results of spectral bandwidth broadening, but without proof of comb operation or retained coherence, have been shown for operation at cryogenic temperatures in earlier studies[12,14,23]. The fact that the coherence in the additionally generated sections of the spectra is not lower than that towards the center, fits well with the explanation of amplified four-wave-mixing (FWM) within the cavity[23]. Through this process new sidebands are generated at the distance corresponding to the injected RF frequency. Compared to free-running, the noise floor of the SWIFTS spectrum is noticeably lower for the RF injection cases. This can be interpreted as a stabilization of the beatnote without impacting the overall coherence.

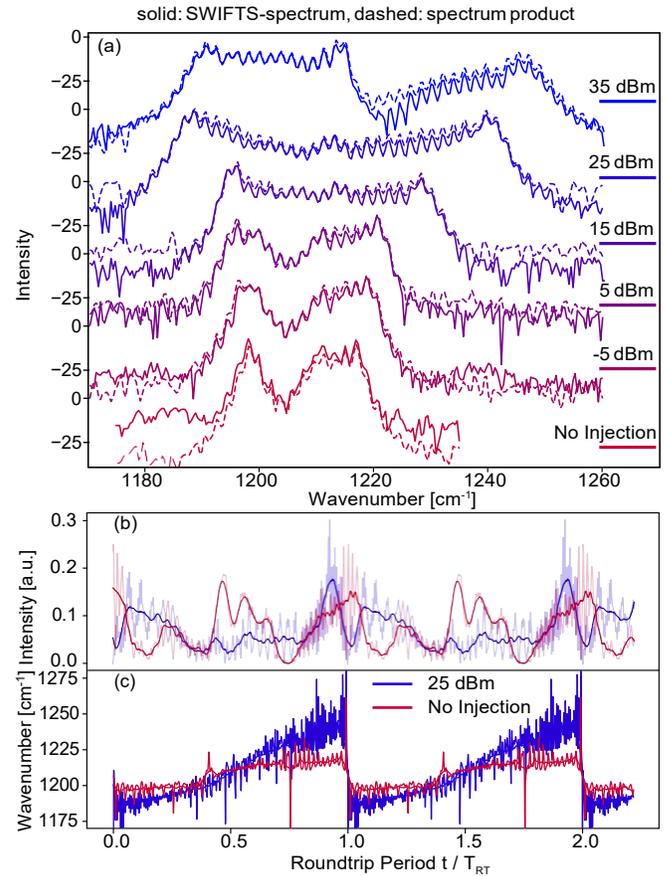

FIG. 3. Results of SWIFTS measurements at 950 mA DC bias and -25°C: (a) Injection power dependent mode intensities of SWIFTS signal and the spectrum product derived from the DC-signal measured with the MCT detector. (b) Reconstructed instantaneous intensity and (c) instantaneous frequency of laser output with no and 25 dBm injection power. The solid lines are the result of treating the raw data with a smoothing filter.

In Fig. 3 (b) & (c) the reconstructed instantaneous intensity and frequency are shown for the free-running (red) and for 25 dBm (blue) injection power for two repetition periods in the time domain. The solid lines are the result of treating the raw data with a smoothing filter. In free-running, the reconstructed time trace during a round-trip period can roughly be split into three lobes, whereas at 25 dBm injection the intensity modulation during a round-trip is closer to having only one main intensity-maximum per round-trip, showing an overall amplitude modulation. The instantaneous frequency of the free-running device has pronounced steps where the frequency jumps between the two spectral high-power sections during one round-trip period. For the RF injection case, these sections have been washed out and correspondingly, the instantaneous frequency is much closer to a linear chirp than in the free-running.

## IV. CONCLUSION - OUTLOOK

The presented work demonstrated that a microstrip-like line waveguide geometry for mid-infrared QCL combs, can indeed enhance the comb properties, in terms of an extended locking range under RF injection, and a broadened optical spectrum, without detriment to the comb coherence. The fact that monolithically grown heavily doped semiconductors are used, instead of wafer bonding on metals, leads to more compact devices and improved performance, with operation closer to room temperature, in contrast to previous attempts of similar designs.

In order to further improve these devices and make them even more effective for practical applications, such as portable dual-comb spectroscopy, there are is a number of steps that can be taken: narrower active regions, as well as a packaging scheme which can allow the epitaxial side down mounting of the device combined with an efficient RF matched circuitry, can surely lead to even better performance, and a more compact solution that wouldn't require an RF probe. In addition, this kind of device, combined with separated passive or active sections for RF modulation, could be excellent candidates for active mode-locking applications.

**FUNDING INFORMATION**

The authors gratefully acknowledge financial support from the BRIDGE program, funded by the Swiss National Science Foundation and Innosuisse, under the project CombTrace (20B2-1_176584/1).

Financial support from the Qombs Project funded by the European Union's Horizon 2020 research and innovation program under Grant Agreement No. 820419 is also gratefully acknowledged.

**SUPPLEMENTARY MATERIAL**

See supplementary material for additional information.

**DATA AVAILABILITY**

The data that support the findings of this study are available from the corresponding author upon reasonable request.

**ACKNOWLEDGMENTS**

The authors would like to thank the staff as well as the equipment responsibles of FIRST - Center for Micro- and Nanoscience cleanroom at ETH Zürich for the use of facilities and continued assistance

The authors would like to thank Zhixin Wang for proof reading the manuscript.